# The macroeconomic effect of the information and communication technology in Hungary

Peter Sasvari
Institute of Business Sciences, Faculty of Economics
University of Miskolc
Miskolc, Hungary

*Abstract*— It was not until the beginning of the 1990s that the effects of information and communication technology on economic growth as well as on the profitability of enterprises raised the interest of researchers. After giving a general description on the relationship between a more intense use of ICT devices and dynamic economic growth, the author identified and explained those four channels that had a robust influence on economic growth and productivity. When comparing the use of information technonology devices in developed as well as in developing countries, the author highlighted the importance of the available additional human capital and the elimination of organizational inflexibilities in the attempt of narrowing the productivity gap between the developed and developing nations. By processing a large quantitiy of information gained from Hungarian enterprises operating in several economic sectors, the author made an attempt to find a strong correlation between the development level of using ICT devices and profitability together with total factor productivity. Although the impact of using ICT devices cannot be measured unequivocally at the microeconomic level because of certain statistical and methodological imperfections, by applying such analytical methods as cluster analysis and correlation and regression calculation, the author managed to prove that both the correlation coefficient and the gradient of the regression trend line showed a positive relationship between the extensive use of information and communication technology and the profitability of enterprises.

## I. INTRODUCTION

Despite its fast expansion and more widespread usage, the effect of information and communication technology on productivity was negligible in the 1980s as well as in the beginning of the 1990s [1]. Later, more and more studies were published arguing that information and communication technology contribute to the growth of productivity.

The effects of ICT on economic growth got into the centre of interest due to the technological revolution taking place in the US in the second half of the 1990s. The previous international productivity trends were altered in the period 1996-2000. The pace of productivity growth, which is defined by production per worker or rather production per labour hour, accelerated in the US and, in contrast to the previous decades, exceeded the growth dynamism of the European Union including its economic and monetary union. With the bursting of the dotcom bubble (the dramatic decrease in the value of blue chip stocks), the analysis of the macro- and microeconomic effects of ICT was overshadowed. Part of the reason of this moderate attention was that considerable excess capacity had become available in the blue chip sector and the real economy needed some time to make use of it. In the period of the recession, or rather of the low growth of GDP, following the bursting of the technological bubble, companies were forced to cut back on their investments in information and communication technology. The acceleration of the growth of productivity made the pace of GDP growth more dynamic as well.

According to the European Committee's report, the macroeconomic effects of information and communication technology predominate in four channels, having an influence on economic growth and productivity [7].

1. The production of information and communication technologies is inseparable from rapid technological development,

2. The investment channel, that is, the growth of production potential by accumulating information and communication technology capital,

3. The possible production externalities (network externalities, network of external economic effects) connected to information and communication technology,

4. The increased demand for information and communication technology may intensify the demand for other types of labour force and capital. However, it must be taken into consideration that ICT is a substitution for other outputs, and the application of new technologies with the necessary restructuralization leads to tensions in the labour and capital markets.

This is the reason why negative growth effects should be taken into account in the case of the fourth channel, at least in the short and medium term. In many cases, information technology may soften the demand for human resources because ICT products themselves can substitute the living work or embodied human capital used to produce them. The

`The described work was carried out as part of the TÁMOP-4.2.1.B-10/2/KONV-2010-0001 project in the framework of the New Hungarian Development Plan. The realization of this project is supported by the European Union, co-financed by the European Social Fund.

substitution of human resources with information and communication technologies will probably increase in the future as technologies and delivery systems become more mature.

In theory, the owners of information and communication technologies can benefit from the information revolution by gaining higher profits, employees can achieve higher salaries or wages, and users can also feel the beneficial effects in the form of lower prices. Based on empirical experiences, profits and salaries showed a modest increase but these changes were slight in comparison to the drop in the relative price of ICT products. This allows us to conclude that the countries employing information technologies benefited slightly more from the information technology revolution than ICT-producing countries as part of their profits were lost due to the unfavourable changes in purchasing power parity. It must be noted that salaries and wages paid in the ICT sector are higher in a great number of developing countries manufacturing information technologies than elsewhere [3].

For the time being, the effect of information technologies is not unambiguous on the geographical location of production in the relation between centre and periphery. On the one hand, as transaction and communication costs become lower, the flexibility of production comes to the front against economy of scale opportunities, a robust increase in the expansion of economic activities; a kind of deconcentration can be expected. On the other hand, the more precise information available on the changes of consumer preferences, the growing weight of intermediate goods used as inputs in the production of other products, and the opportunity of outsourcing certain economic activities to suppliers all make it advantageous for companies to produce close to markets. If outsourcing results in a growing number of new intermediaries providing various partly customized services (accounting, marketing, purchasing etc.), being close to markets is also essential as concerned companies are able to save up more time [4].

The use of information technology products continues to expand rapidly in the developed countries but productivity advantages appear more slowly than in the case of developing countries. Similar to developed nations, the reducing price of information technology products is likely to be the main driving force in the developing world. The use of information technology can be detected in the growth of productivity only if the additional human capital is available, the deregulation of telecommunication infrastructure and information flow takes place and it is possible to eliminate organizational inflexibilities that prevent companies from exploiting the advantages of new technologies and ideas. Although information technologies contribute to raising the standard of productivity in the developing world, the productivity gap may widen between the developed and developing countries.

The high standard of human capital shows a strong correlation with the adaptation of information technologies. As new technologies usually appear in the form of new equipment, high investment rates accelerate their adaptation.

Finally, a strong connection could be observed between the expansion of information technologies and economic policy. The probability of the expansion of new technologies is higher when economic policies are open to imports and the inflow of foreign working capital.

The rate of Internet users and the number of mobile phones is not necessarily lower in some of the poorest countries than in the countries representing a higher economic development level. This allows us to conclude that there is a strong intention to have access to information technologies and international knowledge networks even in the poorer countries. The real question is whether these technologies can be used for accelerating economic growth in those nations.

Information technologies provide attractive opportunities to "by-pass" and exceed out-of-date technologies.

Modern technologies also bring education much faster to a considerably larger number of people than before. With having better access to information and lower transaction costs, people living in the periphery of domestic and international markets can join the mainstream by using up-to-date information technologies. The opportunities of applying information technologies are outstanding in raising productivity, including plants, banks, ports and even governments. These trends are strengthened by continuous innovations and cost reductions.

The effect of information technologies on productivity is not perceptible in the whole developing world. In many cases, the main reasons for this are the lack of adequate complementary human capital, the low responsiveness of the telecommunication sector and a fair amount of inflexibility. Regarding human capital, information technologies may moderate the demand for human resources as IT products themselves substitute the living labour needed to manufacture them. At the same time, the need for complementary human capital in information technologies can be significant, especially in the field of business and government applications.

## II. THE METHOD OF THE RESEARCH

The examination of the subject is interdisciplinary as it has social and scientific references, so a complex approach was needed when I started processing the literature. I needed to study literature on economics, law, sociology and technology connected to the information society.

In consideration of the complexity of the studied subject, I selected several analytical methods and approaches. During the data collection, I reclined upon the Hungarian and the international literature on the subject, thus I was able to process a large quantity of information (nearly 6000 figures). I also extended my literature research to printed and electronic publications on the Internet. As part of my research, I conducted an empirical survey among Hungarian companies and enterprises. The questionnaire was mainly answered by senior directors of the related companies (executive directors, Human Resources managers etc.), in the case of sole proprietorship, sole proprietors themselves as self-employed persons gave the answers to the questions. The questionnaire was filled in by 536 respondents altogether. The sampling unit consisted of Hungarian enterprises operating in several economic sectors; the chosen sampling method was accidental sampling. The applied methods used for processing the primary data of the research were a correlation and regression calculation, multiple regression models and a customized indicator system in SPSS 16.0 [9].

### III. THE MACROECONOMIC EFFECT OF THE INFORMATION AND COMMUNICATION TECHNOLOGY

ICT devices contribute to the improvement of productivity, the economic growth or the acceleration of the economy in several areas. As far as macroeconomic effects are concerned, the technological development is very rapid alongside with the productivity and the total factor productivity (TFP) in the economic sectors producing ICT devices. On the one hand, this process increases the national average in itself, especially when its share tends to grow in the GDP; on the other hand it makes other economic sectors more dynamic by the technological and economic links throughout the whole economic system.

Profits gained with the help of the rapid technological development and the improvement of productivity was eroded by the dropping ICT prices. Countries producing ICT devices lost a part of their profits realized from production because of the deteriorating swap ratio.

The source of productivity and growth benefits from capital deepening (it describes an economy where the amount of capital per worker is increasing), that is the growing rate of using ICT devices, which is stimulated by the huge decrease in ICT prices. These benefits appear in the form of the increased output of existing products and services, manufacturing new products or providing new services, fulfilling customer needs more efficiently and decreasing transition costs etc. As the effect of ICT devices on increased productivity and more dynamic growth are connected to capital deepening, it can be seen that the countries and businesses using these new technologies have benefited more from the revolution of information technology, than the countries producing them.

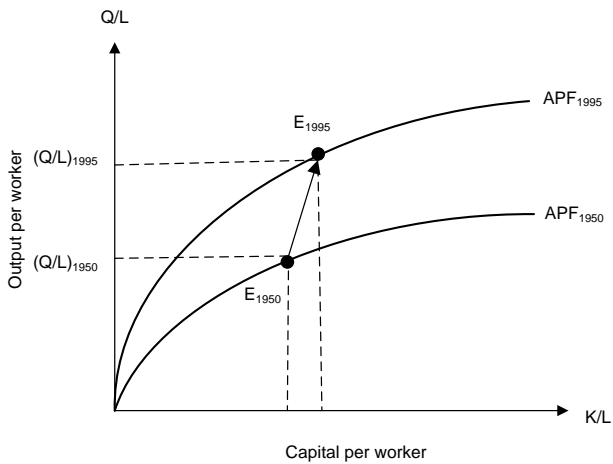

Figure 1. Technological development elongates the production curve [6]

ICT devices also increase the total factor productivity that is they improve the degree of utilization of capital and labour force. The total factor productivity (TFP) is applied to express the overall effect on the savings of economies of scale, management skills, production externalities and other, non-traditional factors influencing productivity. The significance of the growing total factor productivity is that it accelerates the pace of economic growth without any additional costs as well as without having to increase the quantity input. Capital deepening is a necessary but not sufficient condition for improving productivity. It can only unfold in its fullest form when the potential efficiency surplus of ICT devices is exploited. A more dynamic TFP automatically accelerates the pace of labour productivity, thus it helps to boost economic performance.

Using ICT devices also improves productivity and makes economic growth more dynamic because information technology cannot be regarded as capital goods in the traditional sense of the word. The installation of a new information technology device raises the value of other existing devices as well. As a consequence, network effects may occur within companies, moreover they may appear between industrial sectors, and they may necessitate shaping new forms of cooperation (outsourcing).

As it was stated above, ICT devices increase productivity and output by capital deepening, improved total factor productivity and network externalities at the microeconomic level. The advantages of using ICT devices at the macroeconomic level come from all the advantages of the companies' improved productivity and from the network advantages based on the feature of reducing transition costs and accelerating innovation. The network advantage does not depend on the operation of a given company and its business strategy.

However, the effects of ICT devices on the productivity of companies cannot be measured unequivocally at the microeconomic level because of certain statistical and methodological imperfections, the difficulties in measuring network effect at a business level and the lack of data enabling to make international comparisons. Furthermore, the effects of ICT devices on productivity appear at a later time, as they are preceded by a longer or shorter learning process. The productivity paradox has started to vanish by now. It has become clear that statistics cannot or just partially show the secondary effects of using ICT devices in the economy (faster information processing, improvement of productivity in producing knowledge, for instance).

In countries where competition is fierce in the market, enterprises using ICT devices are not necessarily the main winners of capital deepening, it is the customers who can benefit from it by getting lower prices, better quality or more convenience.

It is not necessarily true in countries where competition is weak. Here, companies are able to realize a greater part of benefits coming from capital deepening. But it has its own price as the secondary effects of using ICT devices are more limited in the economy.

With the help of the compound indicator and the financial data of the studied economic sectors, an attempt was made in the research to find a connection between the development levels of ICT and their profitability. Profitability and productivity are influenced by a lot of other factors as well. As it was not possible to measure and show the effect of those other factors, the results are not full but informative.

Based on the statistical connection between the compound indicator and the increment of the Gross Value Added per worker, the correlation coefficient is 0.13, while the gradient of the regression trend line is 0.17. Both numbers show a positive connection between the compound indicator and profitability [8].

Then, using a coordinate system, the connection between the changes of the specific indicators of the studied economic sectors and the development level of those sectors was illustrated. The Y axis shows the growth pace of Gross Value Added per capita in the economic activities between 2003 and 2008 [10]. The X axis shows the compound indicator that was created for measurement purposes. The points defined by the two values show clearly where a given economic sector can be found in the coordinate system, what groups can be constituted, and what tendency can be observed.

The highest increment of specific Gross Value Added was produced by the sectors 'Manufacturing', 'Electricity, gas and water supply', 'Transport, storage and communication' and 'Financial intermediation'. With the exception of 'Electricity, gas and water supply', all of these economic activities belong to the group of underdeveloped sectors (below 50%).

High (but still not reaching the developed status) compound indicators were shown by the sectors 'Mining and quarrying' and 'Wholesale and retail trade; repair work', as they produced an increment of Gross Value Added below the average, these economic sectors can be found in the lower right part of the coordinate system. The sectors 'Construction', 'Health and social work' and 'Hotels and restaurants' can be seen as laggards, so they got into the lower left part of the coordinate system.

The 'Agriculture, hunting and forestry' sector can also be classified as a laggard economic activity, but as the effect of the compound indicator on the increment of Gross Value Added was less significant, it can be found in the upper left part of the coordinate system. Drawing a trend line on the points, it is clear that the line shows a positive gradient, that is, the higher the usage of ICT devices, the higher improvement can be detected in the specific Gross Value Added.

IV. Determining the Net Income of Enterprises by using Multiple Linear Regression including the other variables

The connection between several socioeconomic phenomena is so complicated that the change of a given variable cannot be characterized sufficiently in each case with the help of another variable related to the former one. The analysis has to be extended to many criteria even when the aim of the analysis is only to understand the connection between two criteria, as it is very rare in the economy when the link between two phenomena can be studied separate from other essential effects [5].

The specific methods of studying multiple connections are correlation analysis and multiple regression analysis. The former method measures the strength of the arithmetic relationship between two variables, while the aim of the latter method is to find a standard pattern in stochastic relationships.

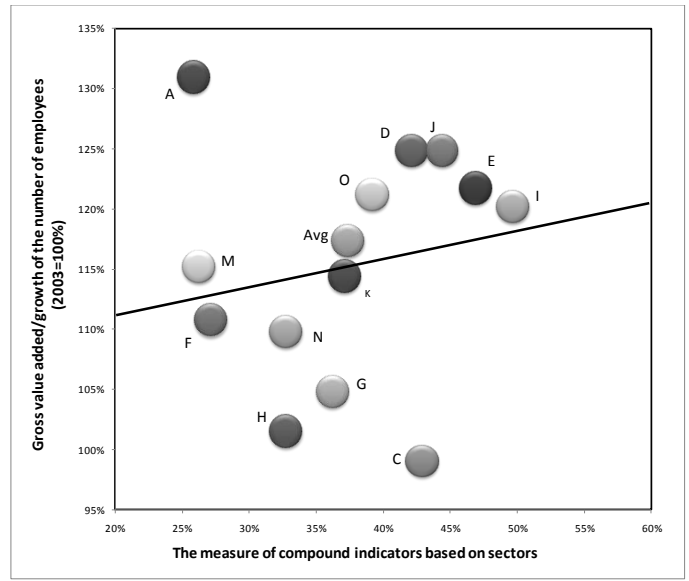

Figure 2. Connection between the growth of gross value added and the development level of information and communication technology in several economic activities[1]

Correlation analysis describes the strength and direction of a linear relationship between two or more variables. In the correlation analysis presented below, I used metric variables.

In my primary research, I expressed the variable of the net income of enterprises with the linear combination of the following continuous variables:

headcount data that are further divided into the number of IT specialists employed, the total number of employees, the number of regular personal computer users, the number of employees using computers connected to the Internet; and the e-commerce financial data such as the value of purchases via the Internet, the sales revenues via the Internet, the value of purchases through computer networks and the sales revenues through computer networks expressed in thousand forints.

| Independent variables | Regression analysis | Dependent variable |
|---|---|---|
| **1. Headcount data:**<br>1.1. The number of IT specialists employed<br>1.2. The total number of employed<br>1.3. The number of regular personal computer users<br>1.4. The number of employees using computers connected to the Internet<br>**2. E-commerce financial data:**<br>2.1. The value of purchases via the Internet<br>2.2. Sales revenues via the Internet<br>2.3. The value of purchases through computer networks<br>2.4. Sales revenues through computer networks | Multiple linear regression | Net sales income of enterprises |

Figure 3. Net income of enterprises expressed with multiple linear regression [5]

The table below shows that seven independent variables got into the multiple regression models with the exception of the sales revenues via the Internet.

---

[1] A=Agriculture, hunting and forestry, C=Mining and quarrying, D=Manufacturing, E=Electricity, gas and water supply, F=Construction, G=Wholesale and retail trade; repair work, H=Hotels and restaurants, I= Transport, storage and communication, J=Financial intermediation, K=Real estate, renting and business activities, M=Education, N=Health and social work, O=Other community, social and personal service activities.

TABLE I.   IDENTIFICATION OF VARIABLES ENTERED OR REMOVED INVOLVED IN THE RESEARCH

| Model | Variables Entered | Variables Removed | Method |
|---|---|---|---|
| 1 | 1.1. The number of IT specialists employed<br>1.2. The total number of employed<br>1.3. The number of regular personal computer users<br>1.4. The number of employees using computers connected to the Internet<br>2.1. The value of purchases via the Internet<br>2.3. The value of purchases through computer networks<br>2.4. Sales revenues through computer networks | | Enter |

Performing a multiple linear regression analysis makes sense only when the stochastic relationship can be demonstrated between the independent variables and the dependent variable. It can be seen in the table that the resulting multiple linear regression model is good because the value of $R^2$ statistics is 0.999 which means that the given model explains 99.9% of all variances.

TABLE II.   SUMMARY OF MULTIPLE LINEAR REGRESSION ANALYSIS (MODEL SUMMARY)

| Model | R | R Square | Adjusted R Square | Std. Error of the Estimate |
|---|---|---|---|---|
| 1 | 1,000 | ,999 | ,998 | 34069,514 |

The analysis of variance shows the significance of the part explained by the applied linear model. In other words it shows whether a model has an explanatory power or not. The second column represents the sum of squares. Total row of the table presents the total sum of squares. The Residual row of the table refers to the partial sum of squares. The third column shows the degrees of freedom. The F-test in the analysis of variance is used to assess whether the expected values of a quantitative variable within several pre-defined groups differ from each other. It serves as a measure of the statistical importance or significance of the differences among the group means. Mean square is the quotient of the sum of squares and degrees of freedom, the values of mean squares are shown in the fourth column. The F-test can be found in the next column, its value determines the value of significance shown in the sixth column. If the value of significance is low (less than 0.005), I have to leave the null hypothesis according to which the independent variables and the dependent variable have no relationship with one another.

It can be proven that the independent variable explains a significant part of the dependent variable, so it is worth studying the strength of their relationship, in other words, the size of the explained part.

As the null hypothesis proved to be true, the result of the following t-test shows that the variable "the number of IT specialists employed" has no significance in the model because its significance level is high.

TABLE III.   REGRESSION COEFFICIENTS IN THE CASE OF USING ENTER METHOD (COEFFICIENTS)

| Model | | Unstandardized Coefficients | | Standardized Coefficients | t | Sig. |
|---|---|---|---|---|---|---|
| | | B | Std. Error | Beta | | |
| 1 | (Constant) | -74879,4 | 15734,4 | | -4,7 | ,018 |
| | 1.3. The number of regular personal computer users | 76222,8 | 6852,5 | 4,6 | 11,1 | ,002 |
| | 2.1. The value of purchases via the Internet | 1,108 | ,120 | ,363 | 9,2 | ,003 |
| | 2.3. The value of purchases through computer networks | 43,73 | 6,157 | 1,813 | 7,1 | ,006 |
| | 2.4. Sales revenues through computer networks | -111,13 | 8,872 | -2,973 | -12,5 | ,001 |
| | 1.1. The number of IT specialists employed | 88415,08 | 67336,68 | ,402 | 1,3 | ,281 |
| | 1.2. The total number of employed | 13672,86 | 2183,621 | 1,514 | 6,3 | ,008 |
| | 1.4. The number of employees using computers connected to the Internet | -67251,95 | 4075,924 | -4,124 | -16,5 | ,000 |

Using the backward method, the removal of the independent variables from the model continues until the partial explanation of all remaining variables has significance.

At first, every independent variable appears in the model, however, it can be seen that one of them does not have a significance in the explanation of the net income of enterprises. In the first step, the variable „the number of IT specialists employed" was left out of the model as it produced the smallest t-value in absolute terms. The six remaining variables in the final model proved to be significant.

The next step is the interpretation of Beta values. Its constant value is -79497. This value denotes where the axis of the net income of enterprises is cut by the hyperplane containing the six remaining variables.

This hyperplane cuts the axis of the main income only if the value of the six remaining variables is 0. The strongest effect may be caused by the increase of the number of regular personal computer users on enterprises. If the value of the purchases on the Internet and through computer networks increases by 1000 forints, the net income of enterprises will rise by 51000 forints (2.1. and 2.3.).

Another interesting feature is that the variable 'sales revenues through computer networks' has a negative effect on the net income of enterprises.

TABLE IV. REGRESSION COEFFICIENTS IN THE CASE OF BACKWARD METHOD (COEFFICIENTS)

| Model | | Unstandardized Coefficients | | Standardized Coefficients | t | Sig. |
|---|---|---|---|---|---|---|
| | | B | Std. Error | Beta | | |
| 1 | (Constant) | -74879,4 | 15734,5 | | -4,76 | ,018 |
| | 1.3. The number of regular personal computer users | 76222,8 | 6852,6 | 4,65 | 11,12 | ,002 |
| | 2.1. The value of purchases via the Internet | 1,1 | ,12 | ,36 | 9,23 | ,003 |
| | 2.3. The value of purchases through computer networks | 43,7 | 6,2 | 1,81 | 7,10 | ,006 |
| | 2.4. Sales revenues through computer networks | -111,1 | 8,8 | -2,97 | -12,53 | ,001 |
| | 1.1. The number of IT specialists employed | 88415,1 | 67336,6 | ,402 | 1,31 | ,281 |
| | 1.2. The total number of employed | 13672,9 | 2183,6 | 1,51 | 6,26 | ,008 |
| | 1.4. The number of employees using computers connected to the Internet | -67251,9 | 4075,9 | -4,12 | -16,50 | ,000 |
| 2 | (Constant) | -79497,0 | 16666,8 | | -4,77 | ,009 |
| | 1.3. The number of regular personal computer users | 73207,6 | 7016,4 | 4,46 | 10,43 | ,000 |
| | 2.1. The value of purchases via the Internet | 1,3 | ,07 | ,41 | 17,19 | ,000 |
| | 2.3. The value of purchases through computer networks | 50,2 | 4,03 | 2,08 | 12,45 | ,000 |
| | 2.4. Sales revenues through computer networks | -106,5 | 8,84 | -2,85 | -12,05 | ,000 |
| | 1.2. The total number of employed | 16249,9 | 1040,1 | 1,79 | 15,62 | ,000 |
| | 1.4. The number of employees using computers connected to the Internet | -67639,1 | 4417,9 | -4,14 | -15,31 | ,000 |

The calculation of beta values all variables are entered into the model in a standardized form with 0 mean and with unit square. Beta values indicate which independent variable has a stronger effect or higher significance, in the case of my model the beta value of the variable 'the number of regular personal computer users' has the strongest effect (4460). It also turned out that the variables 'the number of employees using computers connected to the Internet' and 'sales revenues through computer networks' did not increase the net income of enterprises.

The table below contains the variables excluded from the model.

TABLE V. TABLE 1 - EXCLUDED VARIABLES

| Model | | Beta In | t | Sig. | Partial Correlation | Collinearity Statistics Tolerance |
|---|---|---|---|---|---|---|
| 1 | 2.2. Sales revenues via the Internet | ,053 | ,024 | ,983 | ,017 | 6,621E-5 |
| | 2.2. Sales revenues via the Internet | ,918 | 1,149 | ,334 | ,553 | ,000 |
| 2 | 1.1. The number of IT specialists employed | ,402 | 1,313 | ,281 | ,604 | ,002 |

V. CONCLUSIONS AND SUGGESTIONS FOR THE PRACTICAL USE OF RESEARCH FINDINGS

It was shown by the figures that one possible reason for gaining advantages from using ICT devices at the macroeconomic level come from the improved productivity of enterprises and from the network advantages based on the feature of reducing transition costs and accelerating innovation. As a consequence, ICT devices increase productivity and output by capital deepening, improved total factor productivity and network externalities at the microeconomic level. At the same time, it has become clear that statistics cannot or just partially show the secondary effects of using ICT devices in the economy as a whole (for instance, faster information processing, improvement of productivity in producing knowledge.)

However, I was able to show the positive correlation between the use of ICT devices and the profitability of the Hungarian enterprises in the studied economic sectors as both the correlation coefficient and the gradient of the regression trend line referred to a strong connection between the compound indicator and the profitability of enterprises. Although the end conclusion of the research is that the use of information and communication technology in the studied Hungarian economic sectors can still be regarded as underdeveloped, with the exception of the agricultural sector where the effect of the compound indicator was the least significant, the positive gradient of the trend line showed that the use of ICT devices resulted in a higher improvement in the specific Gross Value Added in almost every Hungarian economic sector.

As it can be predicted that information and communication technology will have an even deeper impact on the operation and profitability of enteprises in the near future, further researches should be conducted in order to measure its scope and extent.

REFERENCES

[1] M. Castells, "The Information Age," Gondolat- Infonia, 2000.
[2] A. Kápolnai, A. Nemeslaki, and R. Pataki, "eBusiness stratégia vállalati felsővezetőknek (E-business strategies for senior management)," Aula, 2002.
[3] Z. L. Karvalics, "Információ, társadalom, történelem, Válogatott írások, (Information, society, history, Selected works)," Typotex Kiadó, 2003.

AUTHOR PROFILE

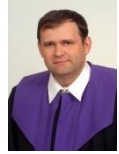

Dr. Peter Sasvari received his MSc in Mechanical Engineering, MSc in Economics and his PhD in Business and Organisation Sciences at the University of Miskolc. Now he is an associate professor at the Institute of Business Science, Faculty of Economics, University of Miskolc, Hungary. His current research interests include different aspects of Economics, ICT and the information society.